\documentclass[preprint,aps]{revtex4-1}
\usepackage{mathrsfs}
\usepackage{gensymb}
\usepackage{amsmath}
\usepackage{amssymb}
\usepackage{bm}
\usepackage{gensymb}
\usepackage{graphicx}
\usepackage{mathrsfs}
\usepackage{multirow} 
\usepackage{subfigure}
\usepackage{textcomp}
\usepackage{units}
\usepackage{natbib}
\usepackage{xcolor}
\definecolor{ddred}{rgb}{0.8,0.5,0.5}

\newcommand{\moire}[0]{moir\'e\ }

\begin{document}

\title{Broadband Neutron Interferometer}
\author{Dmitry A. Pushin}
\email{dmitry.pushin@uwaterloo.ca}
\affiliation{Department of Physics, University of Waterloo, Waterloo, ON, Canada, N2L3G1}
\altaffiliation[Also at ]{Institute for Quantum Computing, University of Waterloo,  Waterloo, ON, Canada, N2L3G1}

\author{Dusan Sarenac} 
\affiliation{Department of Physics, University of Waterloo, Waterloo, ON, Canada, N2L3G1}
\altaffiliation[Also at ]{Institute for Quantum Computing, University of Waterloo,  Waterloo, ON, Canada, N2L3G1}
\author{Dan Hussey}
\affiliation{National Institute of Standards and Technology, Gaithersburg, Maryland 20899, USA}

\author{Houxun Miao}
\affiliation{Biophysics and Biochemistry Center, National Heart, Lung and Blood Institute, National Institutes of Health, Bethesda, Maryland USA}

\author{Muhammad Arif}
\affiliation{National Institute of Standards and Technology, Gaithersburg, Maryland 20899, USA}
\author{David G. Cory}
\affiliation{Institute for Quantum Computing, University of Waterloo,  Waterloo, ON, Canada, N2L3G1}
\altaffiliation[Also at ]{Department of Chemistry, University of Waterloo, Waterloo, ON, Canada, N2L3G1}
\altaffiliation[Also at ]{Perimeter Institute for Theoretical Physics, Waterloo, ON, Canada, N2L2Y5}
\altaffiliation[Also at ]{Canadian Institute for Advanced Research, Toronto, Ontario, Canada, M5G1Z8}

\author{Michael G. Huber}
\affiliation{National Institute of Standards and Technology, Gaithersburg, Maryland 20899, USA}
\author{David Jacobson}
\affiliation{National Institute of Standards and Technology, Gaithersburg, Maryland 20899, USA}
\author{Jacob LaManna}
\affiliation{National Institute of Standards and Technology, Gaithersburg, Maryland 20899, USA}
\author{Joseph D. Parker}
\affiliation{Research and Development Division, Neutron Science and Technology Center, Comprehensive Research Organization for Science and Society (CROSS), 162-1 Shirakata, Tokai, Ibaraki 319-1106 Japan}
\author{Takenao Shinohara}
\affiliation{Materials and Life Science Division, J-PARC Center
Japan Atomic Energy Agency (JAEA), 2-4 Shirakata, Tokai, Ibaraki 319-1195 Japan}
\author{Wakana Ueno}
\affiliation{Materials and Life Science Division, J-PARC Center
Japan Atomic Energy Agency (JAEA), 2-4 Shirakata, Tokai, Ibaraki 319-1195 Japan}

\author{Han Wen}
\affiliation{Biophysics and Biochemistry Center, National Heart, Lung and Blood Institute, National Institutes of Health, Bethesda, Maryland USA}


\begin{abstract}
We demonstrate a two phase-grating, multi-beam neutron interferometer by using a modified Ronchi setup in a far-field regime. The functionality of the interferometer is based on the universal \moire effect that was recently implemented for X-ray phase-contrast imaging in the far-field regime. Interference fringes were achieved with monochromatic, bichromatic, and polychromatic neutron beams; for both continuous and pulsed beams. This far-field neutron interferometry allows for the utilization of the full neutron flux for precise measurements of potential gradients, and expands neutron phase-contrast imaging techniques to more intense polycromatic neutron beams.
\end{abstract}

\date{This manuscript was compiled on \today}

\maketitle

\section*{Introduction}
Interferometers have played a crucial role in the development of science as they allowed for high precision measurements and unique experiments. 
Although early interferometers were based on the interference of electromagnetic waves, the establishment of particle-wave duality expanded the art of interferometry to encompass massive particles as well \cite{Davisson_1927_PR,Estermann_1930_ZfuP,Rauch_1974_PhysLett}. 
The history and recent developments in matter wave interferometry, 
can be found, for example, in these reviews \cite{EIandAI_review,AI_review,Cronin_2009_RMP,Klepp_2014}.

The discovery of the neutron \cite{discovery_neutron} led to the construction of a variety of phase sensitive neutron interferometers. Thermal and cold neutrons are a particularly convenient probe of matter and quantum mechanics given their relatively large mass, nanometer-sized wavelengths, and zero electric charge. One of the first neutron interferometers ~\cite{Maier-Leibnitz1962} was based on wave-front division using a Fresnel biprism setup. The perfect crystal neutron interferometer (NI), based on amplitude division, has achieved the most success due to its size and modest path separation of a few centimeters. Numerous perfect crystal NI experiments have been performed exploring the nature of the neutron and its interactions \cite{ni_book2ed}.
For example, the probing of local gravitational fields \cite{Werner_1988_Physica}, observing the $4\pi$ symmetry of spinor rotation \cite{Rauch_1975_PhysLetta}, observing orbital angular momentum \cite{oam}, putting a limit on the strongly-coupled chameleon field \cite{chameleon}, implementing quantum information algorithms \cite{decoherence}, answering fundamental questions of quantum mechanics \cite{Klepp_2014,Hasegawa2003,Denkmayr2014}, and the precision measurements of coherent and incoherent scattering lengths \cite{Schoen_2003_PhysRev,Huffman_2004_PhysRev,Huber_2014_PRC}. However, perfect crystal neutron interferometry requires extreme forms of environmental isolation \cite{Arif_1994_VibrMonitCont}, which significantly limits its expansion and development.

The advances in micro- and nano-fabrication of periodic structures with features ranging from 1-100 $\mu$m now permit absorption and phase gratings as practical optical components for neutron beams. The first demonstration of a Mach-Zehnder based grating NI in 1985 \cite{Ioffe1985b} used 21 $\mu$m periodic reflection gratings as beam spliters for monochromatic ($\lambda=0.315$ nm) neutrons. A few years later, three transmission phase grating Mach-Zehnder NI was demonstrated for cold neutrons ($0.2$ nm $ <\lambda <  50 $ nm) \cite{GRUBER1989363,VanDerZouw2000,Schellhorn1997} 
with mechanical and holographic gratings \cite{Klepp2011}. The need for cold- or very cold- neutrons with a high degree of collimation limits the use of such interferometers in material science and condensed matter research. An alternative approach was the Talbot-Lau interferometer (TLI) proposed by Clauser and Li for cold potassium ions and x-ray interferometry \cite{Clauser1994}, and implemented by Pfeiffer et al. for neutrons \cite{Pfeiffer2006}. The TLI is based on the near-field Talbot effect \cite{TalbotEffect} and uses a combination of absorption and phase gratings.  In this setup the sample, introduced in front of the phase grating (middle grating), modifies the phase of the transmitted flux through the absorption analyzer grating. While the previously mentioned Mach-Zehnder type grating interferometers  are sensitive to phase shifts induced by a sample located in one arm of the interferometer, this near-field TLI is sensitive to phase gradients caused by a sample. Even though chromatic sensitivity of the TLI is reduced, thus leading to a gain in neutron intensity, in this setup the absorption gratings are challenging to make and curtail the flux reaching the detector; while the neutron wavelength spread will cause contrast loss as the distance to interference fringes (fractional Talbot distance) is inversely proportional to the neutron wavelength.

Here we implement a far-field regime interferometer and report the first demonstration of a multi-beam, broadband NI using exclusively phase gratings. The demonstration includes the use of a continuous monochromatic beam ($\lambda=0.22$ nm), a continuous bichromatic beam ($1/3$ intensity $\lambda_1=0.22$ nm and $2/3$ intensity $\lambda_2=0.44$ nm), a continuous polychromatic beam (approximately given by a Maxwell-Boltzmann distribution with $T_c=40$ K or $\lambda_c=0.5$ nm), and a pulsed neutron beam ($\lambda=0.5$ nm to $\lambda=3.5$ nm). This NI setup consisting of two phase gratings resembles the modified Ronchi setup \cite{Hariharan1974} and functions similarly to the recently demonstrated universal \moire pattern for X-rays and visible light \cite{miao2016universal}.  The advantages of this setup include the use of widely available thermal and cold neutron beams, relaxed grating fabrication and alignment requirements, and broad wavelength acceptance.

\section*{Interferometer Setup}

\begin{figure}
\centering
\includegraphics[width=.8\linewidth]{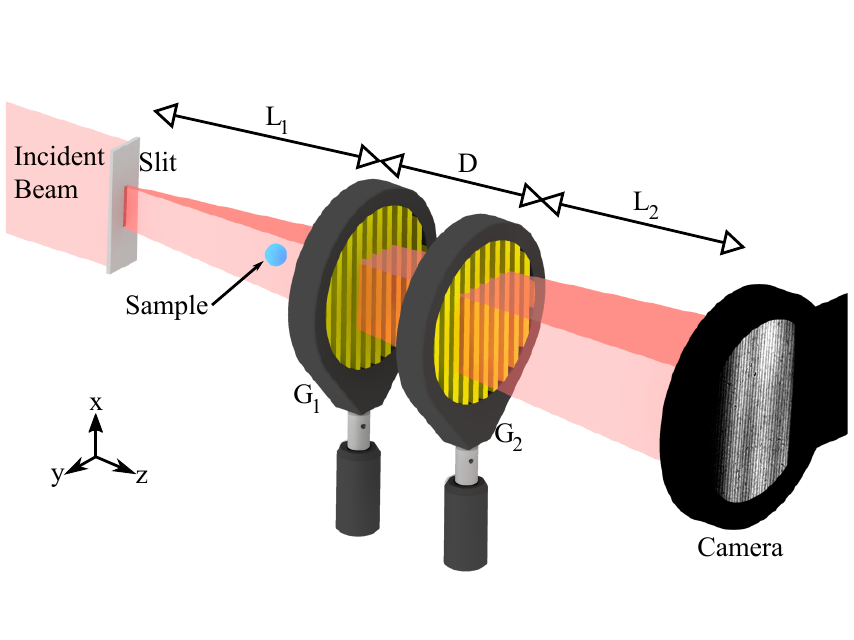}
\caption{Schematic of the two phase-grating interferometer setup. A neutron beam is passed through a narrow slit to define the neutron coherence length along the direction which is perpendicular to the grating fringes. Two identical phase-gratings (G$_1$ and G$_2$) are separated by distance D.  They are placed at a distance L$_1$ from the slit and L$_2$ from the imaging camera. The fringe pattern at the camera is wavelength independent and the fringe visibility can be optimized with the conditions discussed in the text. A sample to be imaged may be placed before the gratings (upstream position) or after the gratings (downstream position).}
\label{fig:setup}
\end{figure}

The experimental setup, shown in Fig.~\ref{fig:setup}, consists of a slit, two identical linear phase gratings of silicon combs with a period of $P_{G_1}=P_{G_2}=2.4$ $\mu$m, and a neutron imaging detector (neutron camera). Although $\pi/2$ phase gratings for $\lambda_{peak}$ give optimal fringe visibility, there were available to us five different gratings with various depths. 

Since the neutron can be described as a matter-wave with a de Brogile wavelength $\lambda=2\pi \hbar/(m_n v_n)$, where $\hbar$ is the reduced Plank's constant, $m_n$ is the neutron mass, and $v_n$ is the neutron velocity, the problem could be treated similar to the X-ray case. The full mathematical treatment of the general situation of polychromatic beam passing through double phase grating setup (see Fig.~\ref{fig:setup}) is described by Miao et al. \cite{miao2016universal}. Here we give a brief description and the key points of the universal \moire effect in a far-field regime for neutrons.

For a majority of this work, a fast neutron produced in a reactor core is first moderated using heavy water to thermal energies and then further cooled using a liquid hydrogen cold source \cite{Williams_2002_Physica} before traversing a neutron guide, the end of which is a slit. After exiting the slit and propagating in free space, the neutron acquires a transverse coherence length of $\ell_c=\lambda \mathscr{L}/s_w$, where $s_w$ is the width of the slit and $\mathscr{L}$ is the distance between the slit and the point of interest. 

In order for the neutron to diffract from the first grating G$_1$ at the distance $L_1$, the neutron's coherence length (along the $y$-axis in Fig.~\ref{fig:setup}), should be at least equal to the period of the grating:
\begin{align}
\ell_c=\frac{\lambda L_1}{s_w} \geqslant P_{G_1}. 
\label{eq:coherence}
\end{align}

The second grating G$_2$ is placed at a distance $D$ from the first grating, and a distance $L_2$ from the neutron camera. 
As neutron cameras have limited spatial resolution $\eta$, the fringe period $P_d$ at the camera should be bigger than neutron camera resolution~\cite{miao2016universal}:
\begin{align}
P_d=\frac{L P_{G_2}}{D} > \eta, 
\end{align}
where $L=L_1+D+L_2$ is the distance between the slit and the camera. Similarly the phase of the fringe pattern on the detector is a periodic function of the slit position, with the period (often called source period) given by~\cite{miao2016universal}:
\begin{align}
P_s=\frac{L P_{G_1}}{D}.
\label{eq:source_period}
\end{align}

Therefore, in order to observe a fringe pattern on the detector the slit width should be smaller than the source period, i.e.
\begin{align*}
P_s < s_w.
\end{align*}
To verify that we are indeed in a far-field regime we consider the Fraunhofer distance when the coherence length is used as the source dimension:
\begin{align}
d_F=2 \frac{\ell_c^2}{\lambda}=2\lambda\Big(\frac{L_1}{s_w}\Big)^2.
\label{eq:fraunhofer_distance}
\end{align}
We consider the coherence length because it is always equal or greater than the grating period in the setup. To satisfy the far-field regime $L_2$ should be greater than the Fraunhofer distance:

\begin{align}
\frac{d_F}{L_2} \approx \frac{\lambda L}{s_w^2} \ll 1.
\label{eq:far_field}
\end{align}
Given the experimental parameters of the monochromatic beamline $d_F/L_2\approx0.04$, polychromatic beamline $d_F/L_2\approx0.02$ for $\lambda=0.5$ nm, and the beamline at the pulsed source $d_F/L_2\approx0.08$ for $\lambda=0.35$ nm. The other two conditions for far-field regime are:
\begin{align}
L_2 \gg \lambda, \quad   
L_2 \gg  s_w. 
\end{align}
In our cases they are the least strict conditions.

The intensity of the fringe pattern recorded by the camera can be fitted to a cosine function
\begin{align}
I=A+B\cos(fx+\phi).
\label{eq:sine}
\end{align}

Thus the mean A, the amplitude B, the frequency f, and the phase $\phi$ can be extracted from the fit. The figure of merit is {\it contrast} or {\it fringe visibility}, which is given by\footnote{Note that eq.~\ref{eq:contrast} is equivalent to $2H_1/H_0$ of reference \cite{miao2016universal}}:
\begin{align}
\mathcal{C}=\frac{\max\{I\}-\min\{I\}}{\max\{I\}+\min\{I\}}=\frac{B}{A}. 
\label{eq:contrast}
\end{align}

If we consider equal period $P_{G_1}=P_{G_2}\equiv P_g$, $\pi/2$-phase gratings, with 50~\% comb fraction, then the maximum contrast is optimized for the condition $\delta_1(\lambda)=\delta_2(\lambda)=0.5$ \cite{miao2016universal}, where:
\begin{align}
\delta_1(\lambda)=\frac{\lambda L_1 D}{L P_g^2}, \quad   
\delta_2(\lambda)=\frac{\lambda L_2 D}{L P_g^2}. 
\end{align}

The closed-form expression of the contrast is given by Eq.12 in \cite{miao2016universal}, and is computed numerically.

\section*{Experimental Methods}

The experiment was performed in four different configurations. The bichromatic and monochromatic beam configurations were performed at the NG7 NIOFa beamline \cite{Shahi2016111} at the National Institute of Standards and Technology Center for Neutron Research (NCNR) with $L_1=1.73$ m and $L=3.52$ m for bichromatic beam and $L_1=1.20$ m and $L=2.99$ m for monochromatic beam. The neutron camera used in this setup has an active area of 25 mm diameter, with scintillator NE426 (ZnS(Ag) type with $^6$Li as neutron converter material) and a spatial resolution of <100 $\mu$m, and virtually no dark current noise  ~\cite{germ_detector}. Images were collected in $300$ s long exposures.
The neutron quantum efficiency of the camera is 18 \% for $\lambda=0.22$ nm and about 50~\% for $\lambda=0.44$ nm. The neutron beam is extracted from a cold neutron guide by a pyrolytic graphite (PG) monochromator with $\lambda=0.44$ nm and $\lambda=0.22$ nm components and approximately 3.2 to 1 ratio in wavelength intensity. To  change from bichromatic to monochromatic configuration, i.e. filter out the $\lambda=0.22$ nm component, a liquid nitrogen cooled  Be-filter with nearly 100 \% filter efficiency ~\cite{Shahi2016111} was installed downstream of the interferometer entrance slit. The slit width was set to $200$ $\mu$m and slit height to $2.5$ cm.

The polychromatic beam configuration was performed at the NG6 Cold Neutron Imaging (CNI) facility \cite{husseybeamline} at the NCNR. The CNI is located on the NG6 end-station and has neutron spectrum approximately given by a Maxwell-Boltzmann distribution with $T_c=40$ K or $\lambda_c=0.5$ nm. The slit to detector length is $L=8.8$ m and slit to G1 distance is $L_1 =4.65$ m. The slit width was set to 500  $\mu$m and slit height to $1.5$ cm. 

The imaging detector is an Andor sCMOS NEO camera viewing a $150~\mu$m thick LiF:ZnS scintillator with a Nikon 85 mm lens with a PK12 extension tube for a reproduction ratio of about 3.7, yielding a spatial resolution of  $\eta=150~\mu$m \footnote{Certain trade names and company products are mentioned in the text or identified in an illustration in order to adequately specify the experimental procedure and equipment used.  In no case does such identification imply recommendation or endorsement by the National Institute of Standards and Technology, nor does it imply that the products are necessarily the best available for the purpose.}. To reduce noise in the sCMOS system, the median of three images were used for analysis. The exposure time was $2$ s for the data in Fig.~\ref{fig:sepdistance} and Fig.~\ref{fig:phasestepping} and $20$s for the data in Fig.~\ref{fig:single_imag} (f)-(h). 

The fourth configuration uses a pulsed neutron beam produced at the Energy-Resolved Neutron Imaging System (RADEN) \cite{RADEN}, located at beam line BL22 of the Japan Proton Accelerator Research Complex (J-PARC) Materials and Life Science Experimental Facility (MLF). The wavelength range that was used was from 0.05 nm to 0.35 nm. The slit to detector length is $L=8.6$ m and slit to G1 distance is $L_1 =4.24$ m. The slit width was set to 200 $\mu$m and slit height to 4 cm. The neutron imaging system employed a micro-pixel chamber ($\mu$PIC), a type of micro-pattern gaseous detector with a two-dimensional strip readout, coupled with an all-digital, high-speed FPGA-based data acquisition system \cite{upic}. This event-type detector records the time-of-arrival of each neutron event relative to the pulse start time for precise measurement of neutron energy, and it has a spatial resolution of $280~\mu$m (FWHM). The readout of the $\mu$PIC detector introduces a fixed-pattern noise structure which is completely removed by normalizing by empty measurements. Thus, the visibility measurements are from open-beam normalized images of the \moire pattern.  The average number of detected neutron events was about 80 per 160 micron pixel with a 4 h integration time.

\section*{Results}

Fig.~\ref{fig:single_imag} (a)-(d) show examples of typical images obtained in optimized configurations for different beamlines: (a) bichromatic beam  with $\lambda_1=0.22$ nm and $\lambda_2=0.44$ nm (b) same beamline as (a) but with a Be-filter in the beam to filter $\lambda_1=0.22$ nm component out (c) polychromatic neutron beam with peak wavelength $\lambda_{c}=0.5$ nm (d) pulsed source for $\lambda=0.25$ nm. In Fig.~\ref{fig:single_imag} (a) the middle dark region corresponds to the collimator which was placed at the front in the setup, and not the grating pattern. As the Be-filter adds divergence to the beam it can be seen that the dark middle region gets washed out in Fig.~\ref{fig:single_imag} (b). The box on each image represents a region of integration along the vertical axis and the integral curve is shown under each image. Such integral curves were used to fit with Eq.~[\ref{eq:sine}] to extract phase, frequency, and compute the contrast via  Eq.~[\ref{eq:contrast}].

\begin{figure}
\includegraphics[width=.8\linewidth]{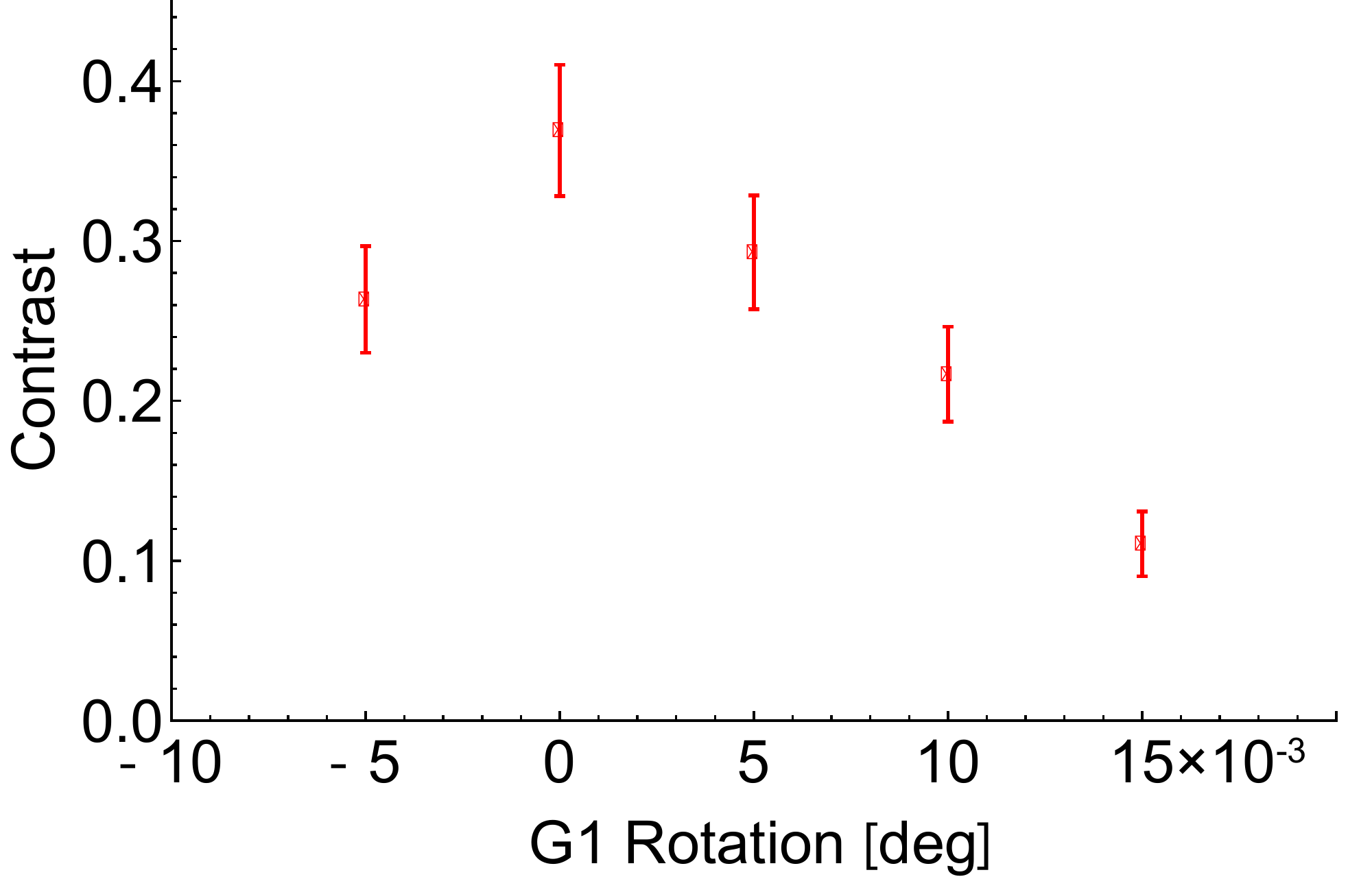}
\includegraphics[width=.8\linewidth]{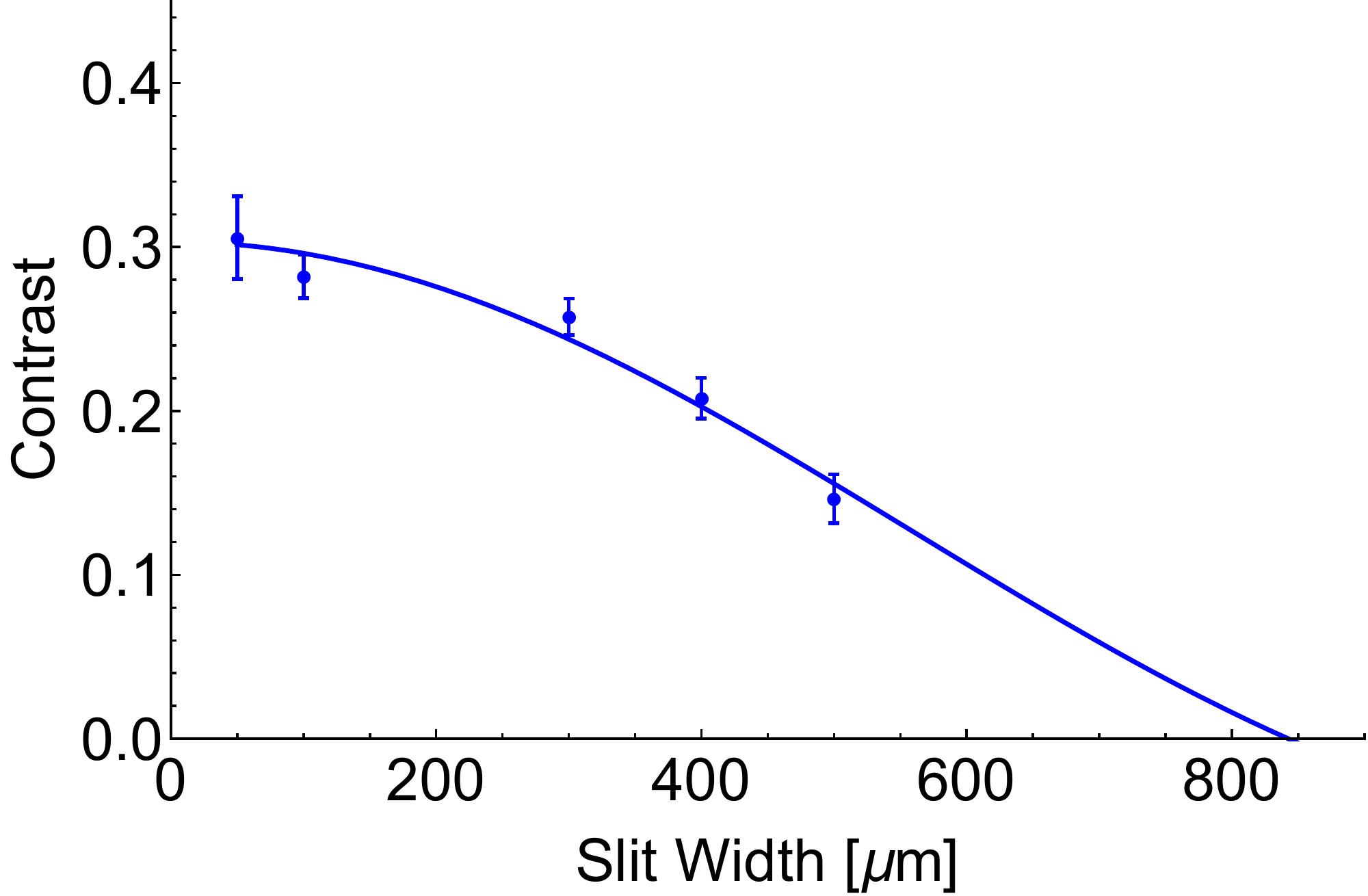}
\caption{Optimization data: (Top) The contrast is found by placing the gratings vertically and rotating one of them along the roll axis in very fine increments. The plot shows the contrast at the monochromatic beamline as a function of the G$_1$  rotation around neutron propagation axis. (Bottom) The dependence of the contrast at the bichromatic setup on the slit width $s_w$. The fit is given by applying Eqn.\ref{eq:contrast_slit} to the data, and shows good agreement with the calculated source period P$_s$.  
}
\label{fig:G1rot}
\end{figure}

To align the gratings the setup is initially arranged with theoretically calculated optimal slit width and lengths $L_1$, $D$, and $L_2$. Then one of the gratings is rotated around the neutron propagation axis ($z$-axis) until the fringe pattern is observed at the camera. The contrast with the monochromatic setup as a function of the first grating rotation around $z$-axis is shown in the top plot of Fig.~\ref{fig:G1rot}. 

The slit height $s_h$ (slit length along the $x$-axis direction in Fig.~\ref{fig:setup}) can be larger than the slit width $s_w$ in order to increase neutron intensity, provided that the gratings are well aligned to be parallel to that direction. The angle range of appreciable contrast is inversely related to the slit height $P_g/(2L_1 s_h/L)$. The expected range of $\pm 0.007\degree$ agrees with the range depicted on Fig.~\ref{fig:G1rot}.

Due to the generally low neutron flux with monochromatic beamlines, the slit widths are optimized for intensity vs contrast. The contrast as a function of the slit width for the bichromatic setup is shown in the bottom plot of Fig.~\ref{fig:G1rot}. Variation of the contrast vs. slit width could be described by the sinc function:
\begin{align}
\mathcal{C}=\mathcal{C}_0 ~\left|\mathrm{sinc}\left[\frac{\pi s}{P_s}\right]\right|.
\label{eq:contrast_slit}
\end{align}
where $\mathcal{C}_0$ is the maximum achievable contrast with a given setup. Thus, given a slit width of one third of the source period, in our case 281 $\mu$m, would give an upper bound of 83 \% contrast. The fit in the bottom plot of Fig.~\ref{fig:G1rot} gives a source period of $P_s=843 \pm 43~\mu$m, which is in good agreement with $P_s=845~\mu$m obtained with Eq.~\ref{eq:source_period} given $D=10$ mm.

\begin{figure}
\includegraphics[width=.8\linewidth]{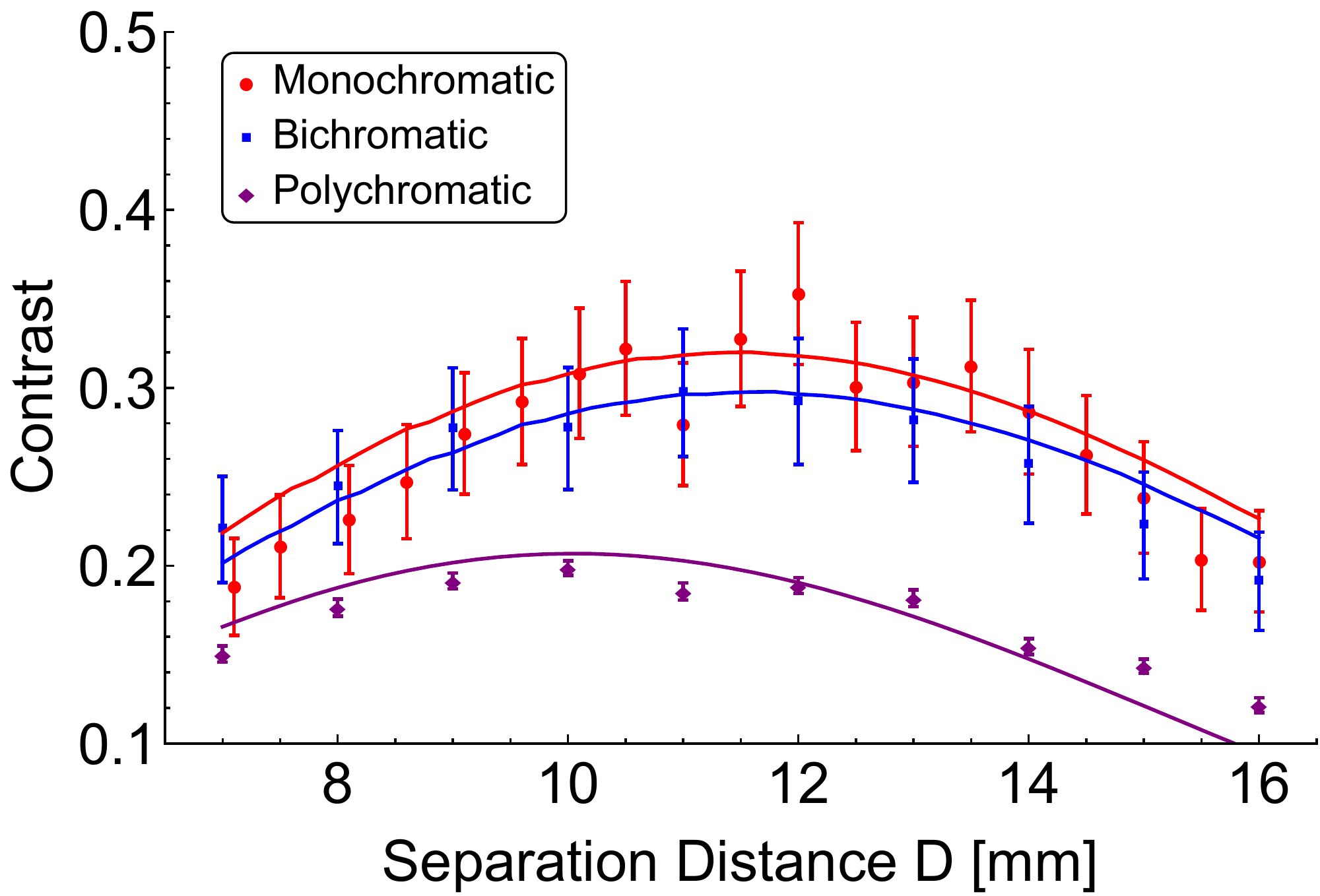}
\includegraphics[width=.8\linewidth]{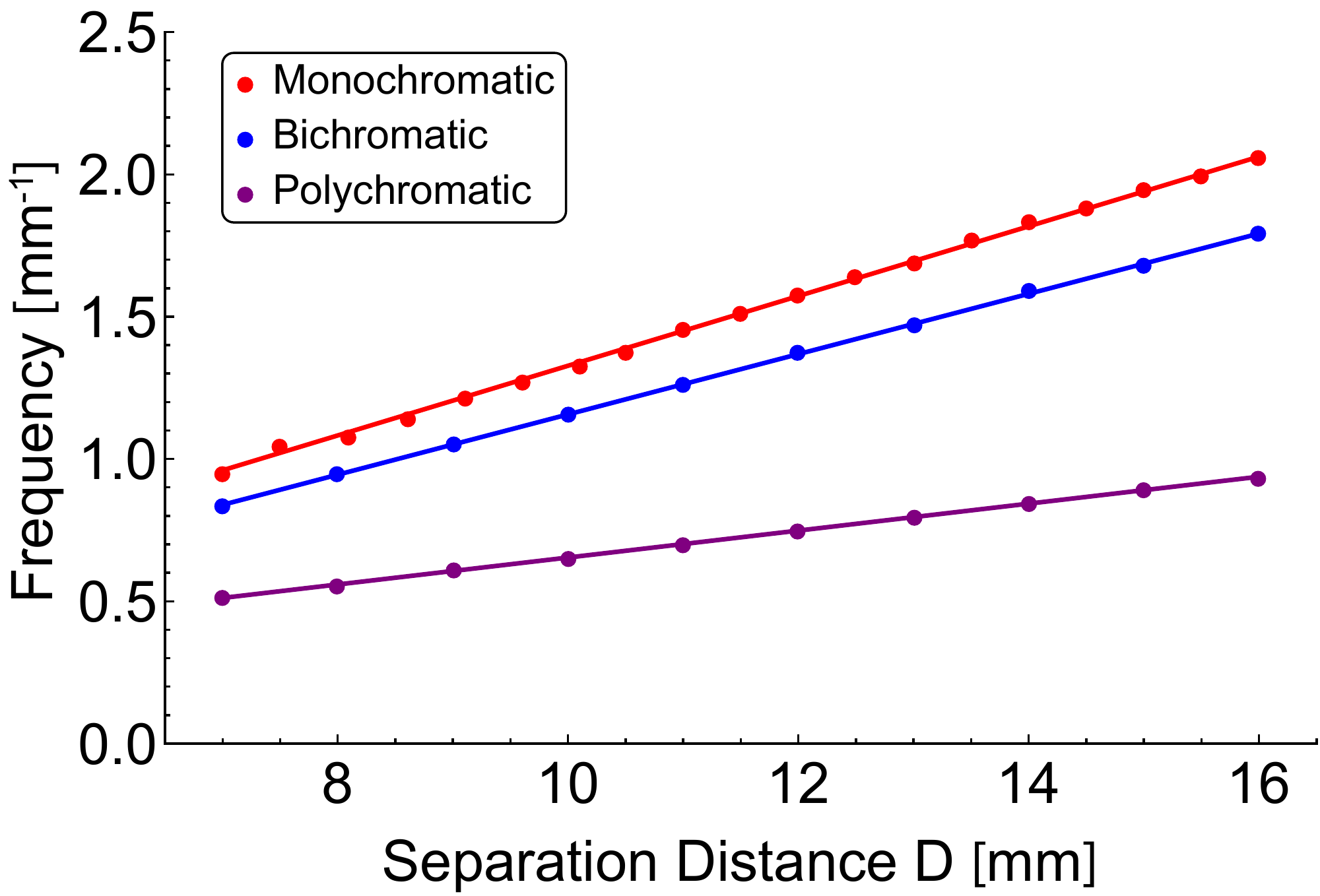}
\caption{(Top) The contrast as a function of the separation distance between the gratings for the monochromatic, bichromatic, and polychromatic neutron beams. The fits are given by Eq.12 in \cite{miao2016universal}, and the contrast for the monochromatic, bichromatic and polychromatic beamlines is optimized at $D=12$ mm, 12 mm, and 10 mm respectively, agreeing well with theoretical predictions \cite{miao2016universal}. (Bottom) The frequency of the oscillation fringes varies linearly as a function of the distance between the gratings. The linear fit is according to $(D-D_0)/(L*P_g)$ which for the monochromatic setup gives $D_0=-0.75$ mm and $L=3.04$ m, the bichromatic setup gives $D_0=-0.8$ mm and $L=3.51$ m, and polychromatic setup gives $D_0=-3.8$ mm and $L=8.36$ m.  
}
\label{fig:sepdistance}
\end{figure}

The top plot in Fig.~\ref{fig:sepdistance} shows contrast (fringe visibility) change versus grating separation, $D$. The data obtained at NIST for the monochromatic, bichromatic and polychromatic beamlines is plotted on the same figure for comparison. The theoretically calculated contrast curves for the three conditions are also plotted, which were based on estimates of $0.27\pi$ phase shift gratings at 0.44 nm wavelength for the mono- and bi-chromatic setups and $0.2\pi$ phase shift at 0.5 nm wavelength for the polychromatic setup. The maximum contrast for the monochromatic, bichromatic and polychromatic beamlines are achieved at $D=12$ mm, 12 mm, and 10 mm respectively, agreeing well with theoretical predictions \cite{miao2016universal}. 
Theoretical estimates indicate that there is room for at least a factor of two improvement of contrast by doubling the grating depths to ~ 17 $\mu$m. The bottom plot in Fig.~\ref{fig:sepdistance} shows the linear dependence of the fringe frequency at the camera on the grating separation. As the distance between the gratings is increased the period of the fringes at the camera is decreased. The linear fit is according to $(D-D_0)/(L*P_g)$ which for the monochromatic setup gives $D_0=-0.75$ mm and $L=3.04$ m, the bichromatic setup gives $D_0=-0.8$ mm and $L=3.51$ m, and polychromatic setup gives $D_0=-3.8$ mm and $L=8.36$ m.  

\begin{figure}
\centering
\includegraphics[width=.6\linewidth]{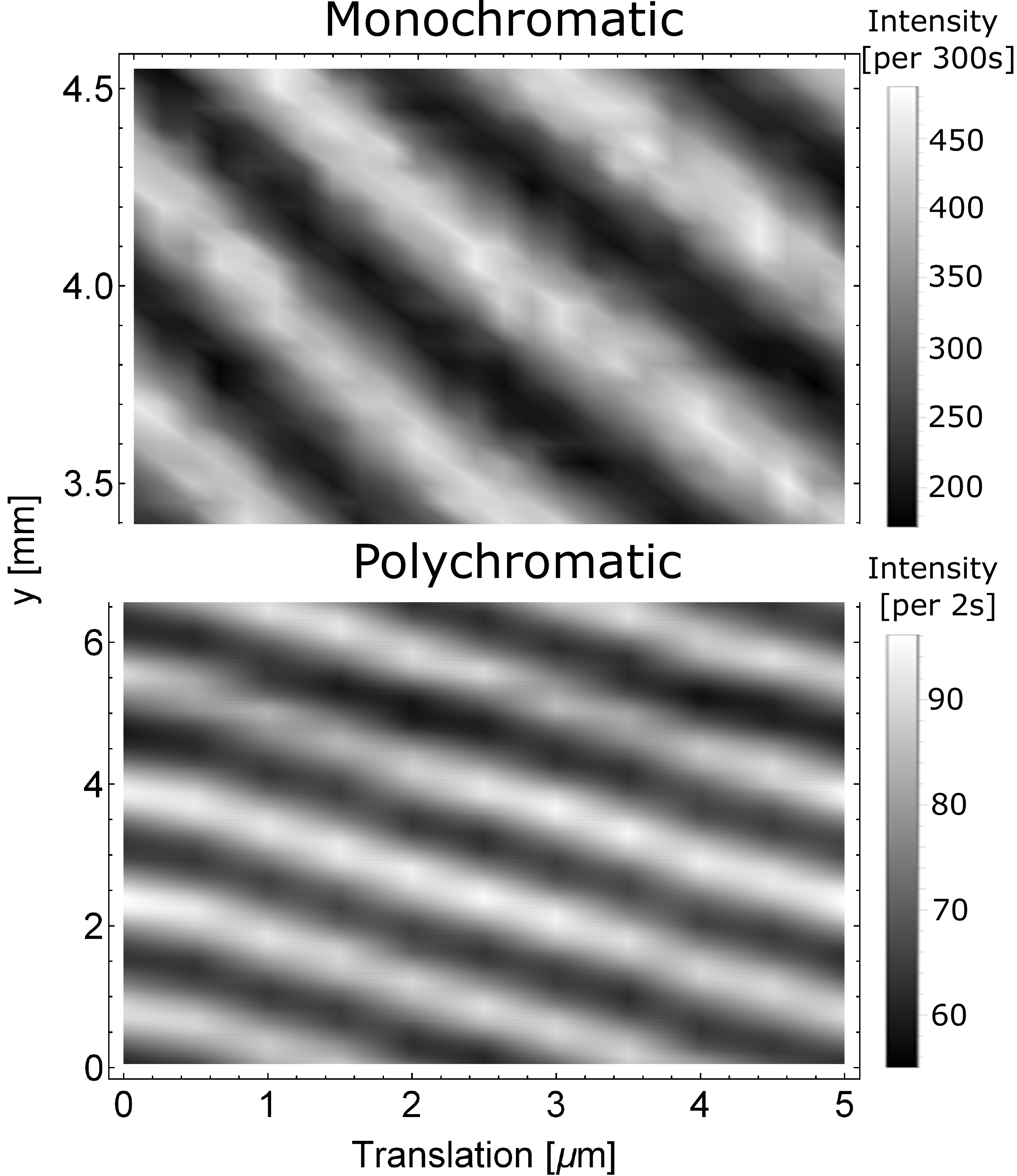}
\caption{Linear phase stepping is achieved with parallel translation of the first grating G1 by increments smaller than the period of the gratings. (Top) Data from the monochromatic beamline (Bottom) Data from the polychromatic beamline. In both cases linear dependence between the phase and grating translation is observed, while the contrast is preserved.}
\label{fig:phasestepping}
\end{figure}

To implement phase stepping of the fringe visibility pattern at the camera one grating needs to be translated in-plane in the perpendicular direction to the grating lines (along the y-axis in Fig.~\ref{fig:setup}). The translation step size needs to be smaller than the grating period. The top plot in Fig.~\ref{fig:phasestepping} shows the 2 dimensional plot of phase stepping for the monochromatic beamline setup; and the bottom plot in Fig.~\ref{fig:phasestepping} shows the phase stepping for the polychromatic beamline setup. In both cases linear dependence between the phase and grating translation is observed, while the contrast is preserved. 

\begin{figure}
\centering
\includegraphics[width=.6\linewidth]{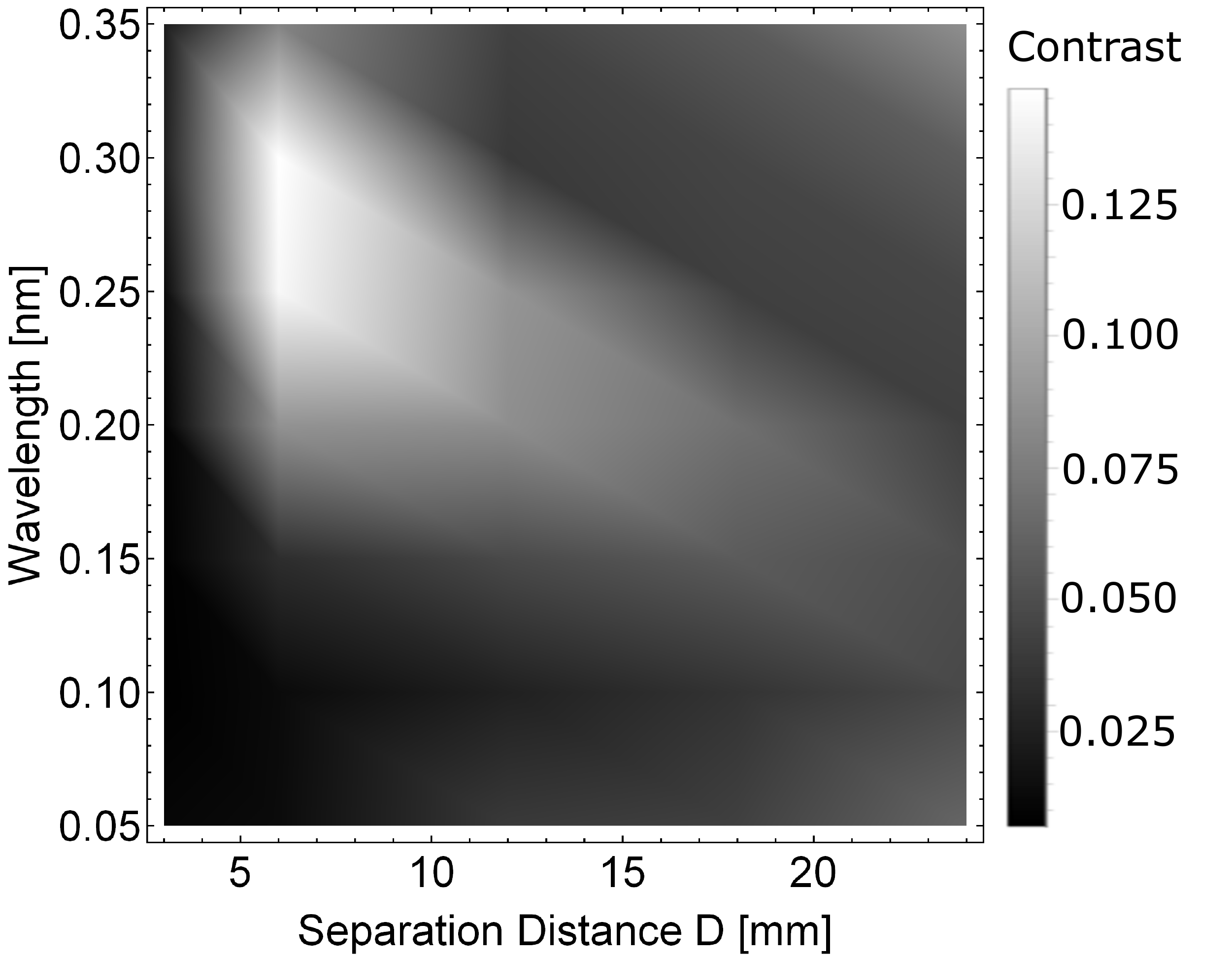}
\caption{Grayscale representation of the contrast (fringe visibility) dependence on neutron wavelength and gratings separation distance, D. Contrast data obtained with a pulsed neutron beam from a spallation source at J-PARC.}
\label{fig:pulsed}
\end{figure}

Similar aligning procedures and measurements were performed at J-PARC spallation pulsed source. Fig.~\ref{fig:pulsed} shows the contrast as a function of the wavelength for various grating separations. Due to the nature of the pulse source we were able to extract contrast as a function of wavelength. Note that at the time of the experiments J-PARC was running at 200 KW as opposed to 1 MW due to technical problems. This lowered the neutron flux to $1/5$ of the standard flux and the low intensity proved to be a significant challenge in terms of optimizing the setup for each independent wavelength. 

\subsection*{Phase-contrast imaging with a polychromatic beam}

\begin{figure}
\centering
\includegraphics[width=.9\linewidth]{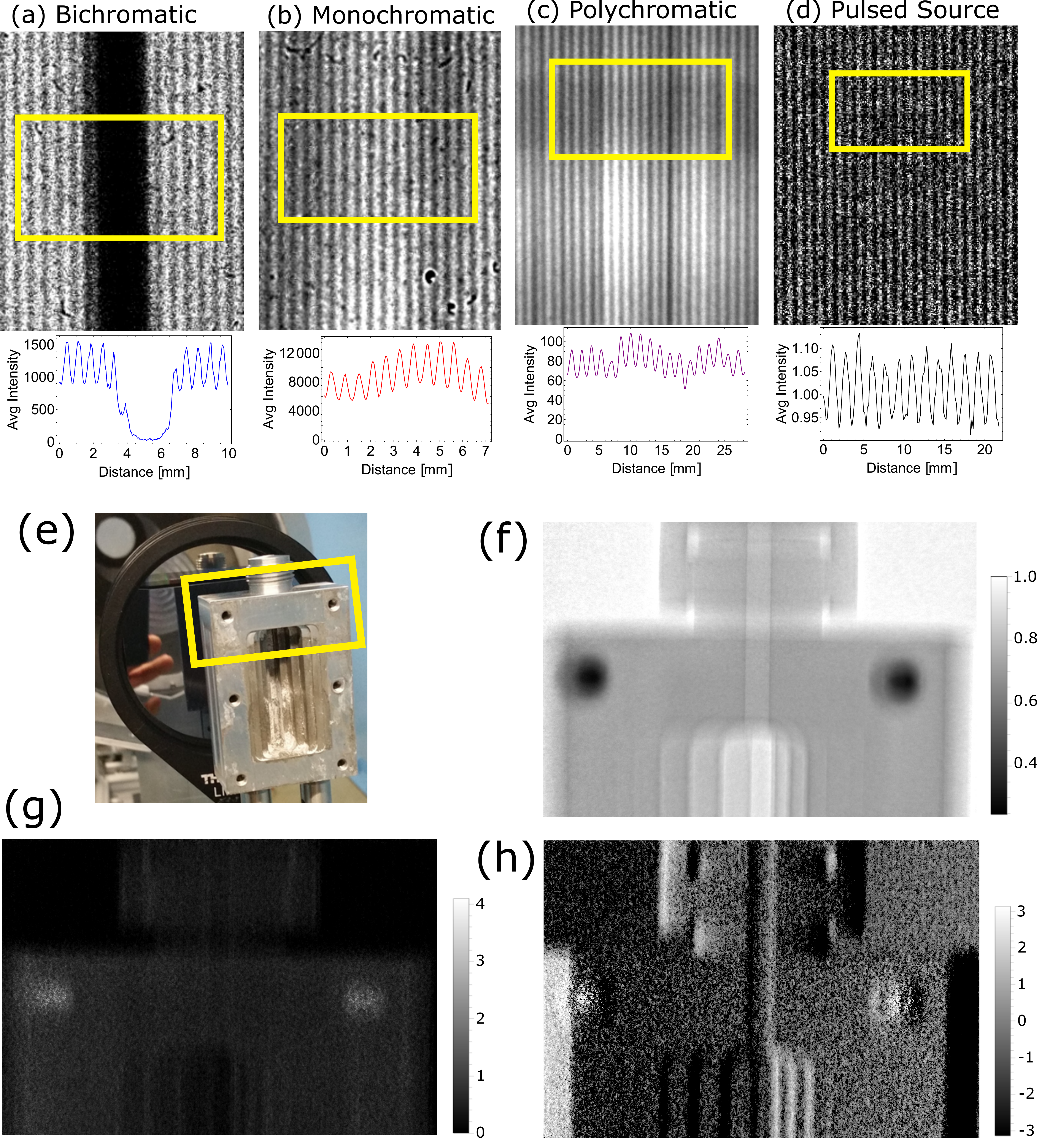}
\caption{{\bf (a)}-{\bf (d)} Typical far-field images obtained with {\bf (a)} 300 s exposure time with the bichromatic neutron beam {\bf (b)} 10000 s exposure time with the monochromatic neutron beam, {\bf (c)} median filter applied on three images with 2 s exposure time with the polychromatic neutron beam {\bf (d)} normalized image from the J-PARC pulsed source. The integrated intensities of the regions specified by the yellow rectangle show the observed fringe pattern at the camera. Note that the dark region in the middle of the bichromatic image is due to the collimator in the particular setup and not due to the gratings. {\bf (e)}-{\bf (h)} Phase-contrast imaging with a polychromatic beam {\bf (e)} The aluminum sample in front of the G$_1$ grating.  The yellow box roughly outlines what is being imaged. The sample has a step profile in the middle region and holes in the corners. {\bf (f)} Neutron attenuation image due to absorption and scattering. Grayscale bar represents transmission through the sample. {\bf (g)}  Spatial variation of the contrast attenuation due to the sample, $-\log(\mathcal{C}_\mathrm{sample}/\mathcal{C}_\mathrm{empty})$. In this case white represents a loss in contrast and black represents no loss in contrast. {\bf (h)} Phase shift in the moir\'e pattern at the detector due to the sample. Grayscale bar represents radians. Here, the white and black patterns represent the highest phase gradient that the neutrons acquire when passing through the sample.
}
\label{fig:single_imag}
\end{figure}

The beam attenuation, decoherence, and phase gradient images shown in Fig.~\ref{fig:single_imag} (f)-(h) are of an aluminum sample shown in  Fig.~\ref{fig:single_imag} (e). The approximate imaged area is depicted by the rectangular box. They images in Fig.~\ref{fig:single_imag} (f)-(h) were obtained by the Fourier transform method described in \cite{fourierphaseimages}. At the described polychromatic beamline at NIST three images with 20 s exposure time were taken at each step in the phase-stepping method  \cite{phasestepping}. A median filter was then applied to every set of three images. The phase step size was 0.24 $\mu$m ranging from 0 to 2.4 $\mu$m of the G$_2$ transverse translation. The G$_1$ - G$_2$ grating separation was $D=11.5$ mm. 

Fig.~\ref{fig:single_imag} (f) shows the conventional attenuation-contrast radiography of the sample, where white color represents full transmission (no attenuation). The shape and features of the sample are well defined in the image. Fig.~\ref{fig:single_imag} (g) shows the decoherence of the fringe contrast due to the sample, $-\log(\mathcal{C}_\mathrm{sample}/\mathcal{C}_\mathrm{empty})$, where the white color represents loss of contrast and the black color represent no contrast reduction. As expected the areas which caused the largest attenuation also caused the largest loss of contrast. Fig.~\ref{fig:single_imag} (h) shows the phase shift in the moir\'e pattern at the detector due to the sample. The white and black patterns represent highest phase gradient that the neutrons acquire when passing through the sample. 

\section*{Conclusion}
For the first time we have demonstrated a functioning two phase-grating based, \moire effect neutron interferometer. The design has a broad wavelength acceptance and requires non-rigorous alignment. The interferometer operates in the far-field regime and can potentially circumvent many limitations of the single crystal and grating based Mach-Zehnder type interferometers, and the near field Talbot-Laue (TL) type interferometers that are in operation today.  Mach-Zehnder type interferometers may provide the most precise and sensitive mode of measurements but a successful implementation requires highly collimated and low energy neutron beams.  On the other hand, a near-field Talbot-Laue interferometer requires absorption gratings and suffers in performance because of a loss in fringe visibility due to the wavelength spread present in a typical neutron beam. 
These constraints limit the wide spread use of these interferometers in a variety of applications. 

The performance of our demonstration interferometer was limited primarily by grating imperfections and misalignments and detector resolution. However, the design is simple and robust without any inherent limitations. We expect that the next generation of interferometers based on the far-field design will open new opportunities in high precision phase based measurements in materials science, condensed matter physics, and bioscience research. In particular, because of the \moire fringe exploitation in this type of interferometers, the uses will be highly suitable for the studies of biological membranes, polymer thin films, and materials structure. Also, the modest cost and the simplicity of assembly and operation will allow this type of interferometers to have wide acceptance in small to modest research reactor facilities worldwide.

\begin{acknowledgments}
This work was supported by the U.S. Department of Commerce, the NIST Radiation and Physics Division, the Director's office of NIST, the NIST Center for Neutron Research, and the National Institute of Standards and Technology (NIST) Quantum Information Program. This work was also supported by the Canadian Excellence Research Chairs (CERC) program, the Natural Sciences and Engineering Research Council of Canada (NSERC) Discovery program, and the Collaborative Research and Training Experience (CREATE) program. D.A.P. is  grateful for discussions with Michael Slutsky.
\end{acknowledgments}
\bibliographystyle{apsrev4-1}  
\bibliography{pnasbib-2}

\newpage

\end{document}